# Signature of Correlated Insulator in Electric Field Controlled Superlattice

Jiacheng Sun[1], Sayed Ali Akbar Ghorashi[1], Kenji Watanabe[2], Takashi Taniguchi[3], Fernando Camino[4], Jennifer Cano[1,5], Xu Du[1]

[1]Department of Physics and Astronomy, Stony Brook University, Stony Brook, New York 11794-3800, USA

[2] Research Center for Electronic and Optical Materials, National Institute for Materials Science, 1-1 Namiki, Tsukuba 305-0044, Japan

[3] Research Center for Materials Nanoarchitectonics, National Institute for Materials Science, 1-1 Namiki, Tsukuba 305-0044, Japan

[4]Center for Functional Nanomaterials, Brookhaven National Laboratory, Upton New York 11973, USA

[5]Center for Computational Quantum Physics, Flatiron Institute, New York, New York 10010, USA.

**The Bloch electron energy spectrum of a crystalline solid is determined by the underlying lattice structure at the atomic level. In a 2-dimensional (2d) crystal it is possible to impose a "superlattice" with nanometer-scale periodicity, allowing to tune the fundamental Bloch electron spectrum, and enabling novel physical properties which are not accessible in the original crystal. In recent years, a top-down approach for creating 2d superlattices on monolayer graphene by means of nanopatterned electric gates has been studied, which allows the formation of isolated energy bands and Hofstadter Butterfly physics in quantizing magnetic fields. Within this approach, however, evidence of electron correlations – which drive many problems at the forefront of physics research – remains to be uncovered. In this work we demonstrate signatures of a correlated insulator phase in**

**Bernal-stacked bilayer graphene (BLG) modulated by a gate-defined superlattice potential, manifested as a set of resistance peaks centered at carrier densities of integer multiples of a single electron per unit cell of the superlattice potential. We associate the correlated insulator phase to the formation of flat energy bands due to the superlattice potential combined with inversion symmetry breaking. Inducing correlated electron phases with nanopatterning defined electric gates paves the way to custom-designed superlattices with arbitrary geometries and symmetries for studying band structure engineering and strongly correlated electrons in 2d materials.**

The electronic properties of crystals originate from the Bloch states described by the quantum mechanical electron wave functions in the presence of the crystal lattice. The Bloch states can also be modified by the presence of a periodic potential superimposed on top of the underlying crystal lattice. A highly celebrated example is the moiré heterostructure, formed by stacking two 2d crystals with a slight mismatch in lattice constants or with a slight rotation. The resulting superlattice (moiré pattern) induces a wide variety of charge transport phenomena, from single-particle-natured satellite resistance peaks and Hofstadter butterfly physics, to the formation of flat energy bands and the associated phases driven by electron correlations, including superconductivity[1-3], Chern insulators[3-7], Mott insulators[8-10], Wigner crystals[11], and orbital ferromagnets[3]. Followed by the experimental discovery of correlated insulators and superconductivity in magic-angle twisted BLG (TBLG), a wide variety of twisted heterostructures have been identified to share similar or partially similar properties, including twisted trilayer graphene[12-14], twisted double BLG[15-18], twisted transition metal dichalcogenides[11, 19, 20], etc. The commonality of these different moiré systems suggests the importance of the superlattice potential in the formation of flat bands and the electron correlations.

Beyond moiré superlattices, a different approach for creating a superlattice potential on a 2d material is by means of the electric field effect using a patterned gate[21-23]. This approach in principle allows much higher flexibility in the geometry and symmetry of the superlattice potentials. Realized through nanopatterning, it bypasses the long-range strain-induced inhomogeneity problem common for the moiré heterostructures. Understanding the impact of such superlattice potential in absence of a moiré pattern provides valuable insight to the origin of some of the exotic behaviors observed in the twisted bilayer systems. Experimentally, effective realization of a gate-induced superlattice potential has been challenging due to the small pitch-size required for the superlattice. The energy scale associated with the superlattice can be estimated to be $\sim hv_F/a$, where $h$ is the Planck's constant, $v_F$ is the Fermi velocity and $a$ is the lattice constant of the superlattice potential. For this energy scale to be sufficiently larger than the random doping induced Fermi energy fluctuations, a pitch size of no more than a few tens of nanometers is required for the superlattice. With the developments of nanolithography, improvements on 2d material device quality, and ultrathin 2d insulators such as hexagonal boron nitride (hBN), recent experimental works have demonstrated 2d[21] and one-dimensional (1d)[22] superlattice potentials through gate dielectric patterning in monolayer graphene samples, with a superlattice periodicity as small as 16nm[23] – comparable to the moiré length in magic angle TBLG. In these devices, the superlattice potential has been shown to induce higher order satellite Dirac points. However, since a Bravais superlattice always preserves two-fold rotational symmetry, such a potential applied to monolayer graphene cannot gap the Dirac cone, which limits its ability to create flat energy bands and achieve correlated electron phases. This is not the case for Bernal-stacked BLG in the presence of a superlattice (SL) potential, which has been recently proposed as a tunable and realistic platform to realize flat energy bands[24]. In this work, we experimentally demonstrate the ability to tune the electronic properties of BLG by a SL potential using the dielectric patterning approach.

The sample studied in this work is a Bernal-stacked BLG encapsulated in hBN layers placed on a SiO$_2$/Si substrate, with the SL structure patterned onto the surface of the SiO$_2$ by dry-etching of a 50 nm-pitch triangular lattice of antidots ~50 nm deep[21] (Figure 1a, b, c). In the presence of a back gate voltage between the BLG and the conducting Si substrate, a periodic electric field contrast is formed as a result of the difference in dielectric constants where the encapsulated BLG sits on SiO$_2$ ($\epsilon_{SiO_2} \approx 4$) versus where it sits over a antidot ($\epsilon_{vacuum} = 1$). Significant SL potential strength can be achieved when the bottom hBN thickness is much smaller than the diameter of the antidot (See Supplementary Information on electrostatic simulations). A metal film top gate on the top hBN allows uniform tuning of the carrier density and Fermi energy throughout the BLG channel. The difference between the top and the bottom gates also induces a displacement field which manifests the differential doping on the two graphene layers. In BLG, the band structure is tuned by both the SL potential and the displacement field, while the Fermi energy is controlled by the overall doping. In this work the tuning knobs are limited to two gate electrodes, allowing access to part of the parameter space with the SL potential and displacement field co-adjusted by the back gate. The top gate tunes the overall Fermi energy as well as the displacement field. To find out the gate dependences of all the potential parameters, we combine the contribution of each gate to each graphene layer, considering an approximate screening efficiency of $\alpha$~0.7: ~70% of the gate induced charge is on the adjacent graphene layer, and ~30% of the charge is on the remote graphene layer[25] (a detailed discussion can be found in the Supplementary Information). The overall periodic charge density modulation due to the back gate-induced SL potential can be described as $n_{SL} = c_{SL}V_{BG}$, and the differential charge density due to the displacement field is $n_{DIS} = (2\alpha - 1)(c_{TG}V_{TG} - \bar{c}_{BG}V_{BG})$, where $c_{SL}$, $\bar{c}_{BG}$ and $c_{TG}$ are the effective capacitances associated with the SL potential, back gate and top gate, respectively. Under small magnitude of $V_{BG}$ where the SL effect is weak, $c_{TG}$ can be precisely determined from measuring the quantum Hall effect while sweeping the top gate voltage $V_{TG}$ at fixed back gate, by correlating $V_{TG}$ at the quantum Hall plateaus with the corresponding filling factor $\nu$ : $c_{TG} = \nu eB/(hV_{TG})$. The mean (averaged) carrier density is $n = c_{TG}V_{TG} + \bar{c}_{BG}V_{BG}$. Measuring the gate-dependence of the primary charge neutral point (defined here as the charge neutral point in absence of a SL potential) at $n_{TOT} = 0$ (i.e., $\frac{c_{TG}}{\bar{c}_{BG}} = -\Delta V_{BG}/\Delta V_{TG}$) allows determination of the effective capacitance ratio between the back and top gates. With both $c_{TG}$ and $\bar{c}_{BG}$ determined, the mean carrier density and the displacement field can then be calculated. The spatial profile of the electric field from the SL patterned SiO$_2$ gate dielectric can be estimated through the electrostatic simulations using COMSOL (see Supplementary Information), from which $c_{SL}$ can be obtained.

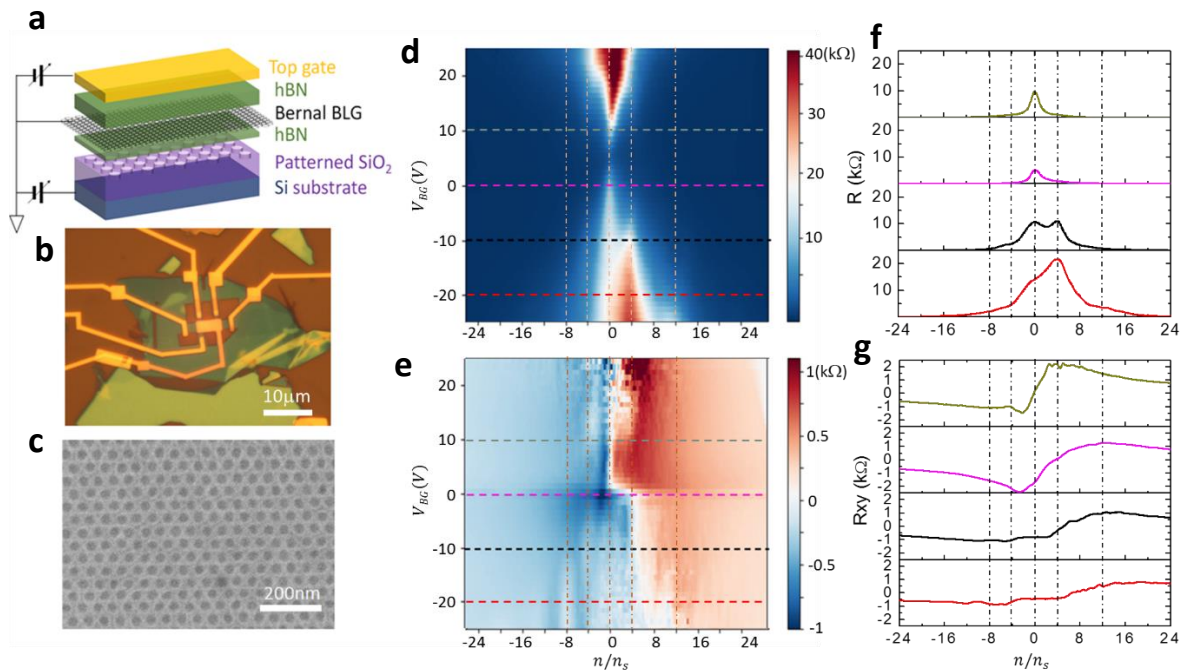

*Figure 1. Dual-gated Bernal-stacked BLG sample and basic characteristics. a. Schematic of the sample structure. b. Optical micrograph of the sample. c. Scanning electron microscope image of the patterned $SiO_2$ surface, showing a triangular lattice with 50nm-pitch. Plots d. and e. represent the color coded zero-magnetic field resistance and low magnetic field Hall resistance as functions of mean carrier density and back gate voltage. Here the averaged carrier density is normalized to the number of electrons per SL unit cell. The Hall resistance measurement was performed in low magnetic field of $\pm 0.5$ Tesla: $[R_{xy}(0.5T) - R_{xy}(-0.5T)]/2$. The vertical dotted lines indicate the mean carrier densities at -8, -4, 0, 4, and 12 electrons per unit cell of the SL potential. The horizontal line-cuts of the plots labeled by the colored dotted lines are presented in f. and g. following the same order and with the same colors, for longitudinal resistance and Hall resistance, respectively. All the measurements were taken at a temperature of 300mK.*

Figure 1d plots the dependence of electrical resistance on the mean carrier density and the back gate voltage. Here the mean carrier density is normalized to the number of electrons per unit cell of the superlattice potential $n/n_s$, where $n_s = 2/(\sqrt{3}a^2)$ corresponds to one electron per unit cell of the superlattice (here $a = 50$ nm is the pitch size). The back gate simultaneously adjusts the SL potential, displacement field strength and the overall doping level. At small back gate voltages, a single resistance peak appears at the primary charge neutral point, as expected for intrinsic Bernal-stacked BLG. Increasing the back gate voltage amplitude, the charge-neutral resistance maximum increases as a result of inversion symmetry breaking from the displacement field, which opens up a band gap [26-28]. At the same time, additional resistance peaks emerge at mean carrier densities equal to integer multiples of $4n_s$, most noticeable at $-4n_s$, $4n_s$, and $12n_s$ (Figure 1d). This is consistent with the formation of energetically isolated superlattice bands with four-fold (spin and valley) degeneracy. We note the absence of a resistance peak at $8n_s$, which suggests that the second and the third superlattice bands from the original charge neutrality are not effectively energetically isolated. As shown in Figure 1d, for negative back gate voltages, the

SL-induced features are significantly more pronounced on the electron side. Similar particle-hole asymmetry has been observed in monolayer graphene under SL potential[21], which can be attributed to the asymmetry in the minima/maxima of the triangular SL potential, i.e., the electrons in the conduction band localize in the potential minima, which form a honeycomb lattice, while the holes in the valence band localize in the potential maxima, which form a triangular lattice. In the case of BLG, additional asymmetry may also come from the gate-dependence of the displacement field. We also observed that the resistance peak associated with the primary neutral point becomes overwhelmed by that from the nearest SL-induced-band gap on the electron side, under moderate negative back gate voltage. The details of these SL-associated features are displayed in Figure 1f which plots the mean carrier density dependent resistance at several back gate voltages, indicated by the colored dotted lines in Figure 1d. The superlattice-induced resistance peaks are significantly more pronounced under negative back gate voltages. As discussed below, this can be explained by the unintentional charge doping which is modulated by the patterned $SiO_2$ surface, enhancing the SL potential strength for $V_{BG} < 0$.

In addition to the longitudinal resistance peaks, Hall resistance in low magnetic field shows oscillatory features in the carrier density dependence which, depending on the back gate voltage, can be noticeable around densities $n/n_s \sim -8, -4, 0, 4, 12$ (Figure 1e). Figure 1g plots the carrier density dependence of Hall resistance at several back gate voltages (example traces indicated by the colored dotted lines in Figure 1e). With increasing back gate voltage amplitude, the transition between the energy-continuums of the electron and hole branches, indicated by the sign flipping around $n = 0$, becomes increasingly broad with low-energy oscillatory features in between. These observations are consistent with charge carrier sign-changes at the longitudinal resistance peaks, hence the formation of band gaps between the SL bands. We note that compared to a moiré SL in twisted bilayer systems with SL period <20nm, the 50 nm-pitch SL samples studied here have a unit cell area which is nearly an order of magnitude larger. Hence the carrier density associated with filling up a non-interacting SL band ($\sim 4.6 \times 10^{10} \text{cm}^{-2}$) is nearly an order of magnitude smaller, and is approaching the random doping fluctuation level of the sample. As a result of the significance of doping fluctuations under such large SL periodicity, the charge transport features associated with the SL bands, both in longitudinal and in Hall resistance, appear strongly bunched and smeared. Nevertheless, all the SL-associated features are reproducible in different samples and within the same sample over thermal cycles.

To further confirm the SL-induced energy bands, we next turn to quantizing magnetic fields. In absence of a SL potential, we expect to observe conventional Landau level sequence associated with the Bernal-stacked BLG. This is observed under moderate positive back gate voltages, where $R_{xx}$ shows resistance valleys in the Landau fan diagram displaying a filling factor sequence of $\nu = 4N$ ($N$ is an integer), as shown in Figure 2a. Under zero back gate voltage, however, we observed additional Landau fan features which originate from the SL bands, besides those which extrapolate to the primary charge neutral point. In Figure 2b, the carrier density and magnetic flux dependence of the $R_{xx}$ valleys are compared to calculations from the Diophantine equation: $\frac{n}{n_s} = t\frac{\phi}{\phi_0} + s$. Here $\phi = B/n_s$ is the magnetic flux through the SL unit cell; $\phi_0 = h/e$ is the magnetic flux quantum; and $t$ and $s$ are integers which correspond to number of electrons per magnetic flux quantum and per SL unit cell, respectively. While the SL-associated Landau fan features are not as clearly distinct as in twisted bilayer systems (again, due to the large SL period limited by the fabrication), we can identify several sets of fan lines computed from the Diophantine equation which are consistent with the data, as shown in Figure

2b. This allows for a two-parameter labeling of the Landau fan features. The Landau fan lines originate from $s = 4N$, indicating the non-interacting nature of the SL bands under weak periodic potential.

The presence of SL bands in zero back gate voltage suggests a "residual" SL potential. This may be due to unintentional charge doping which forms a periodic pattern in accordance with the SL pattern on the surface of the $SiO_2$. This assumption is qualitatively consistent with the observation that the minimum charge neutrality peak resistance, which corresponds to the minimum displacement field at the charge neutrality, happens at a negative back gate voltage of ~−5V (Figure 1d), indicating residue electron doping on the bottom graphene layer. One possible scenario is that the BLG is locally electron-doped at locations of the $SiO_2$ antidots. In presence of a mild positive back gate voltage, the SL induced by the back gate (which induces lower electron density at the locations of the antidots than in between the antidots, due to the dielectric constant contrast) would compensate and smoothen out the SL potential. This explains why the SL potential associated features in the Landau fan diagram disappears under a moderate positive back gate voltage, but appears under zero back gate voltage, as shown in Figure 2.

Under large negative back gate voltages, we expect strong SL potential and displacement field. There, the Landau fan diagram shows a drastically different pattern from that in the weak SL potential. The Landau fans with the conventional BLG sequence ($s = 0$) extrapolate to the origin ($n, B = 0$) which does not coincide with an apparent zero-field resistance peak, while the main resistance maximum corresponds to a charge filling of $\frac{n}{n_s} \sim 4$. Besides these original Landau fans, vertical stripes corresponding to resistance peaks whose fillings do not depend on magnetic field (i.e., $t = 0$) are observed on the electron side of the resistance peak, as shown by the vertical lines in Figure 2c.

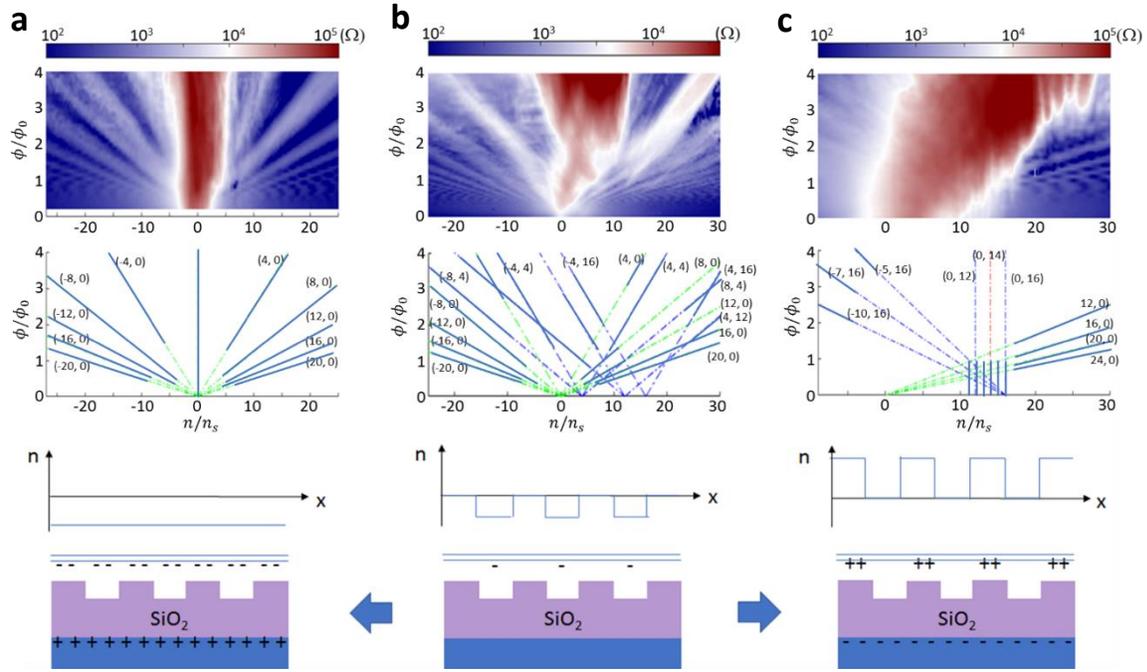

*Figure 2. Landau fan diagrams under different SL potential profiles, at back gate voltages of $V_{BG}$=15V (a), 0V (b) and -30V (c). At zero back gate, Landau fans from Bernal-stacked BLG (s=0) as well as a few side bands (s=4, 8, 12, 16) can be identified. The SL-associated features may be attributed to unintentional doping which is effectively patterned by the SL structure, with BLG locally electron-doped at the locations of the antidots. The unintentional doping profile is flattened under a mild positive back gate voltage, when higher electron density is gate-induced between the antidots compensating the unintentional electron doping at the antidots. This results in the conventional Landau fans for the Bernal-stacked BLG. Under a negative back gate voltage, the SL potential is enhanced giving rise to different Landau fan patterns. For all back gate voltages, the Landau fan lines are calculated from the Diophantine equation and labeled by (t, s). All the measurements were taken at a temperature of 300mK.*

Next, we focus on the strong SL potential regime and discuss the main finding of this work, when the negative back gate voltage is further pushed to larger values. In zero magnetic field, with the back gate voltage beyond ~ -30V, a striking feature becomes evident in the resistance versus charge density and back gate voltage dependence, shown in Figure 3a. A comb-like set of resistance peaks develop at integer multiples of single electron per SL unit cell, most evidently at $\frac{n}{n_s} = 7, 8, 9, ... 14$ at a back gate voltage between -30V and a maximum applied value of -42V. These resistance peaks correspond to not only full-, but also half- and quarter- fillings of the SL-associated bands, and are signatures of an interaction effect which breaks the four-fold degeneracy of each SL-associated band, resulting in correlated insulating behavior at the Coulomb gaps similar to that observed in magic angle TBLG. Unlike in magic angle TBLG, the filling factors here show a wide range of values which would correspond to a stack of flat energy bands. In Figure 3c, we measured the Landau fan diagram under a back gate voltage of -42V. The comb-like resistance peaks in zero magnetic field are non-dispersive in magnetic field, developing into vertical stripes with changing magnetic field dominating over the fillings up to $s\sim$15-20 (i.e., up to 5 SL bands). At larger densities beyond the regime of the comb-like features, we observe Landau fans with $s = 0$, which trace to $n = 0$.

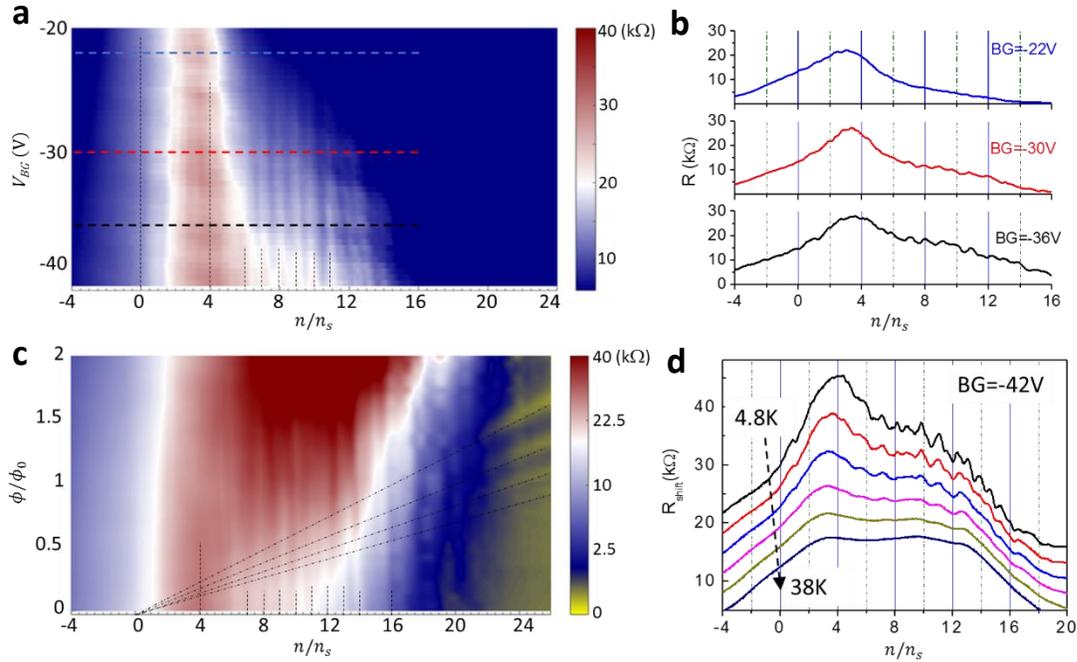

*Figure 3. Signatures of flat bands and correlated insulator under large SL potential. a. Color-coded resistance as a function of normalized carrier density and back voltage. The comb-like stripes which correspond to resistance peaks are highlighted by the dotted lines, situated at integer multiples of single-charge-per-SL-unit-cell. The line-cuts of the plot at colored dashed lines are shown in b. c. Landau fan diagram measured at $V_{BG}$=-42V. At large carrier densities, fan lines associated with Bernal-stacked BLG can be identified which extrapolate to the origin. At lower densities, resistance peaks which do not shift with magnetic field are observed and are highlighted by dotted lines. d. Temperature dependence of the resistance peaks.*

In twisted bilayer systems, the correlated behavior is commonly associated with the formation of flat energy band, with diminishing kinetic energy and diverging density of state. With gate-defined SL potential on BLG, it has been theoretically proposed that a stack of flat bands can form under a large SL potential and displacement field[24]. Our observation of resistance peaks at integer multiples of single-electron-per-unit-cell of the superlattice potential over a wide range of carrier density is consistent with the formation of such a stack of flat bands, with the signature of correlated insulator behavior they are associated with.

Figure 3d plots the temperature dependence of the resistance peaks at all the fillings. Here again we note that while the resistance peak features are weak, the filling factors of the resistant peaks are temperature/thermal cycle independent, which suggest that their origin is not the random disorder-associated universal conductance fluctuations (UCFs). We also note that while UCFs are observed in our samples at the lowest temperatures down to ~0.3K, they become largely absent at above 4K, and have very different back gate voltage dependence compared to the resistance peaks which are stationary at integer fillings of the SL unit cells (a discussion on UCF in our samples can be found in the Supplementary Information.) We argue that resistance peaks at integer fillings are associated with the energy gap between the isolated SL bands when a band is fully filled, as well as the Coulomb energy gap between the interaction-split flat bands when a band is half- or quarter-filled. It is observed that most of the resistance peaks at half- and quarter-fillings become thermally smeared at ~20K. On the other hand, the resistance peaks associated with the non-interacting band gaps (e.g., at fillings of 4 and 12 electrons per SL unit cell) persist up to much higher temperatures.

In order to understand our experimental observations, we model the Bernal-stacked BLG in the presence of a patterned superlattice potential and a displacement field by a four-band continuum model (consisting of two layers and two spins) with an applied 2d cosine potential. Specifically, the Hamiltonian takes the form,

$$H = H_{BLG} + H_{SL} + H_V$$

where the three terms describe the Hamiltonian of intrinsic Bernal-stacked BLG, a spatially varying superlattice potential, and a uniform layer-dependent mean potential, respectively. First, intrinsic Bernal-stacked BLG can be represented as,

$$H_{BLG} = \hbar v_F (\chi k_x \sigma^1 + k_y \sigma^2) + \frac{t}{2}(\tau^1 \sigma^1 - \tau^2 \sigma^2)$$

where $\chi = \pm$ is the valley index, $v_F$ is the Fermi velocity of graphene, $k$ is the electron crystal momentum, and $t$ is the interlayer coupling; Pauli matrices $\tau$ and $\sigma$ correspond to the layer and sublattice spaces. The spatially modulated superlattice potential is described by,

$$H_{SL} = \frac{1}{2}\begin{bmatrix} V_{SL}^{(1)}\sigma^0 & 0_2 \\ 0_2 & V_{SL}^{(2)}\sigma^0 \end{bmatrix} \sum_{n=1}^{6} \cos(Q_n \cdot r)$$

where $V_{SL}^{(1),(2)}$ are the effective strength of the superlattice potential on each layer and $Q_n$ are its wavevectors. For a triangular superlattice potential the mini-Brillouin zone (mBZ) is defined by $Q_n = Q(\cos(\frac{2n\pi}{6}), \sin(\frac{2n\pi}{6}))$ which define the high-symmetry points $\Gamma_m = (0,0), M_m = \frac{1}{2}Q_0$ and $K_m = \frac{1}{3}(Q_0 + Q_1)$. In our calculations, the back gate-induced superlattice potential on each graphene layer is estimated based on the electrostatic simulations of the actual device structure (see Supplementary Information). The superlattice potential is also contributed from an unintentional doping from the back gate side, which can be represented as an effective back gate voltage $V_D \sim 15V$. The resulting superlattice potential follows the empirical relations: $V_{SL}^{(1)} = 0.498(V_{BG} - V_D)$ [meV] and $V_{SL}^{(2)} = 0.213(V_{BG} - V_D)$ [meV].

Finally, the mean potential can be captured by,

$$H_V = \begin{bmatrix} V_0^{(1)}\sigma^0 & 0_2 \\ 0_2 & V_0^{(2)}\sigma^0 \end{bmatrix},$$

with the mean potential on the two graphene layers contributed from the back gate (see Supplementary Information), the top gate and the unintentional doping: $V_0^{(1)} = 1.17 V_T + 2.73 V_{BG} + 1.12 V_D$ [meV] for the bottom graphene layer, and $V_0^{(2)} = 2.73 V_T + 1.17 V_{BG} + 0.48 V_D$ [meV] for the top graphene layer. Here $V_T$ is the equivalent top gate voltage normalized by the back gate capacitance.

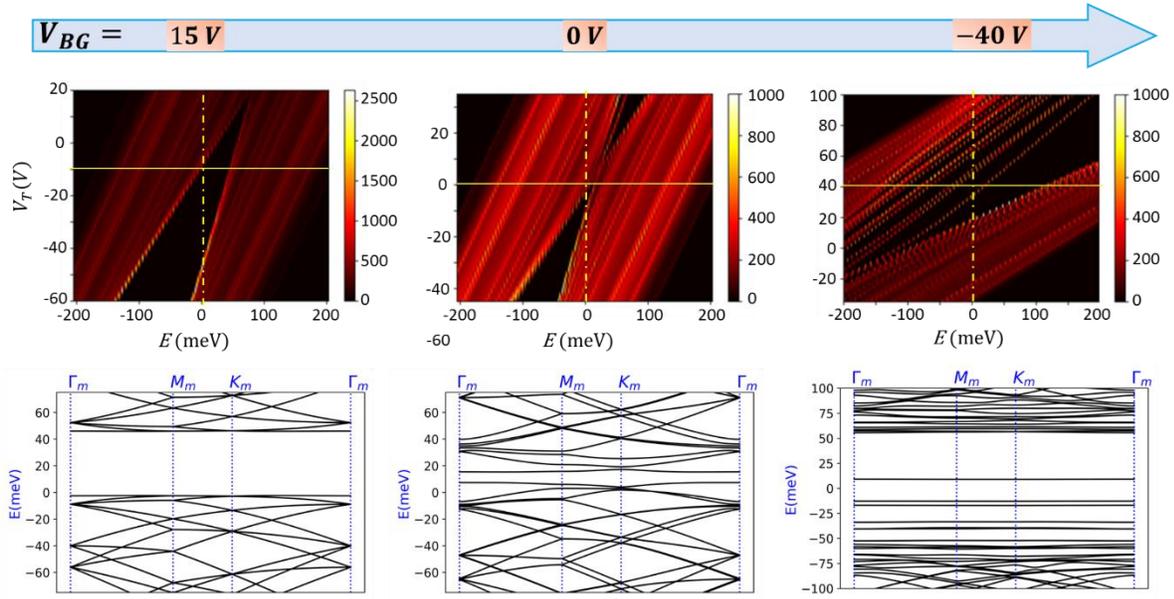

Figure 4. Calculated band structures and DOS at various back gate voltages. The top row shows the color-coded DOS versus energy and $V_T$. The charge transport measurements are related to the DOS at the Fermi level indicated by the vertical dotted yellow lines at E=0. The bottom row shows the band structures at $V_T$ indicated by the solid yellow lines in the top row.

Based on the above parameters, we calculate the band structure of Bernal-stacked BLG under a superlattice potential and its evolution in $V_{BG}$. Figure 4 (top row) shows the density of states (DOS) with respect to $V_T$ and energy for various values of back gate voltages, and (bottom row) the band structure corresponding to the horizontal lines shown in DOS plots. In absence of a superlattice potential, the DOS plot at $V_{BG} = 15\text{V}$ shows an energy gap separating the energy-continuums of conduction and valence bands, as expected for a Bernal-stacked BLG. Changing the back gate voltage towards negative values and hence increasing the superlattice potential strength, multiple flat energy bands emerge at low energies, showing up in the DOS plots as sharp lines which indicate narrow bandwidth. The experimental observations in our transport measurements correspond to zero- energy line-cuts of the DOS plots. In the limit of strong superlattice potential and displacement field, the theoretical calculations show that these flat bands move across with Fermi level with the ramping of $V_T$. Interaction effects can cause these flat bands to break the four-fold degeneracy and split into non-degenerate flat energy bands which are filled by one electron per superlattice unit cell for each band. This is consistent with the aforementioned observations.

Finally, we note that the modeling of the DOS is also qualitatively consistent with the observations in the non-interacting regime under low SL potential. Zooming into the charge neutrality, the evolution of the SL bands under $V_T$ can be traced (see Supplementary Information) from which one can identify that the 2$^{nd}$ and the 3$^{rd}$ SL bands are nearly degenerate, resulting in the absence of a measurable band gap at $\frac{n}{n_s} = 8$. Moreover, increasing the magnitude of $V_{BG}$ at low SL potential and displacement field, the band gap between the 1$^{st}$ and the 2$^{nd}$ SL bands increases faster than the band gap associated with the associated primary neutrality point, resulting in the resistance peak at $\frac{n}{n_s} = 4$ to overwhelm the primary charge neutrality resistance peak.

In summary, we have studied Bernal-stacked BLG under tunable SL potential and displacement field. SL-associated energy bands are observed, manifested by resistance peaks at remote band gaps in zero magnetic field and their corresponding Landau fans in quantizing magnetic field. Under strong SL potential, signatures of correlated insulator phases are observed, manifested as a set of resistance peaks centered at carrier densities of integer multiples of single-electron-per-unit-cell of the superlattice potential. We attribute the correlated electrons to the formation of flat energy bands due to the superlattice potential combined with inversion symmetry breaking. Inducing correlated electron phases with nanopatterning defined electric gates paves the way to custom-designed superlattices with arbitrary geometries and symmetries for studying band structure engineering and strongly correlated electrons in 2d materials.

**Acknowledgement**

X.D acknowledge support from NSF awards under Grant No. DMR-1808491 and DMR-2104781. K.W. and T.T. acknowledge support from the JSPS KAKENHI (Grant Numbers 20H00354, 21H05233 and 23H02052) and World Premier International Research Center Initiative (WPI), MEXT, Japan. S.A.A.G. and J.C. acknowledge support from the Air Force Office of Scientific Research under Grant No. FA9550-20-1-0260 and acknowledge hospitality from the Kavli Institute for Theoretical Physics, where this research was supported in part by the National Science Foundation under Grant No. NSF PHY-1748958. J.C. also acknowledges support from the Flatiron Institute, a division of the Simons Foundation, and the Alfred P. Sloan Foundation through a Sloan Research Fellowship. This research used the Electron Microscopy facility of the Center for Functional Nanomaterials (CFN), which is a U.S. Department of Energy